\documentclass[a4paper,11pt]{article}
\pdfoutput=1 

\usepackage{jinstpub} 

\usepackage{graphicx}
\usepackage{subcaption}

\title{\boldmath Development of a Test System for Data Links of the ATLAS Inner Tracker (ITk) Upgrade Silicon Pixel Detector}

\author[a,1]{F.~Ustuner,\note{Corresponding author.}}
\author[b,c]{A.C.~Mullins,}
\author[a]{S.~Eisenhardt,}
\author[b]{M.~Kocian,}
\author[b]{D.~Su,}
\author[b]{M.~Wittgen,}
\author[b]{A.~Young}

\affiliation[a]{The University of Edinburgh,\\Edinburgh, The United Kingdom}
\affiliation[b]{SLAC National Accelerator Laboratory\\California, The USA}
\affiliation[c]{Southern Methoist University,\\Dallas, The USA}

\emailAdd{fuat.ustuner@cern.ch}

\abstract{This contribution introduces a novel test system developed to evaluate the signal transmission quality in high-speed data links for the 2026 Inner Tracker (ITk) upgrade of the ATLAS experiment. Using an FPGA-based data acquisition (DAQ) framework, the setup can run simultaneous Bit Error Rate (BER) tests for up to 64 channels and generate virtual eye diagrams, for qualifying the $\sim$26k electrical links at the ATLAS ITk data rate of 1.28Gb/s. The paper includes results from system calibration, yielding its contribution to the measured losses, and preliminary results from tests of prototype and pre-production assemblies of on-detector links of the three ATLAS ITk Pixel subsystems.}


\arxivnumber{2503.09303} 

\proceeding{Topical Workshop on Electronics for Particle Physics\\
  30 September- 4 October 2024\\
  Glasgow, The United Kingdom}

\notoc
\begin{document}
\maketitle
\flushbottom

\section{Introduction}

The ATLAS experiment 2026 Inner Tracker (ITk) upgrade will include silicon pixel detectors, requiring high-bandwidth readout. A total of $\sim$26k command and data links, operating at a rate of 1.28Gb/s, will be needed to read out the three ITk subsystems (Outer Endcaps, Outer Barrel and Inner System). The high-irradiation environment requires up to six-meter-long copper links between the front end and the conversion to optical signals. The links must be of the lowest possible mass, radiation hard and match the allocated signal loss budget of 13dB at 640MHz. Custom developments of gauge 34 AWG Twinax cables with polyethene dielectric and aluminium shield have been demonstrated to fulfill the specifications. The signal loss in single links was validated with eye diagrams from fast oscilloscopes, a time-consuming method unsuitable for all ITk links.  This work presents the development of an FPGA-based data acquisition (DAQ) system capable of conducting simultaneous Bit Error Rate (BER) scans, at the ATLAS ITk Pixel data rate of 1.28Gb/s, recording virtual eye diagrams for up to 32 links per connected assembly, and being scalable to test multiple assemblies at the same time.

\section{Eye and BER Diagrams}

For signal transmission with $\ge$1Gb/s the signal loss is the critical parameter when testing the transmission quality. The eye diagram is the standard visualization method used to identify this loss \cite{1}, allowing for an easy examination of the transmission quality of digital signals, as it overlays multiple signal passages, showing the transitions between logic levels 0 and 1 relative to the logic unit cell. From the variation of the transitions one can determine the noise and jitter of the signal as well as the signal loss for a given transfer function of a receiver. 

However, this method is not scalable to test 26k channels. Instead, to perform automated Quality Control (QC) tests on our Twinax data link assemblies, we have developed a method to obtain scans of BER tests, mapping the offsets in time and voltage across the logic unit cell.

\begin{figure} [htbp]
\centering
\subfloat[]{\includegraphics[width=0.5\textwidth]{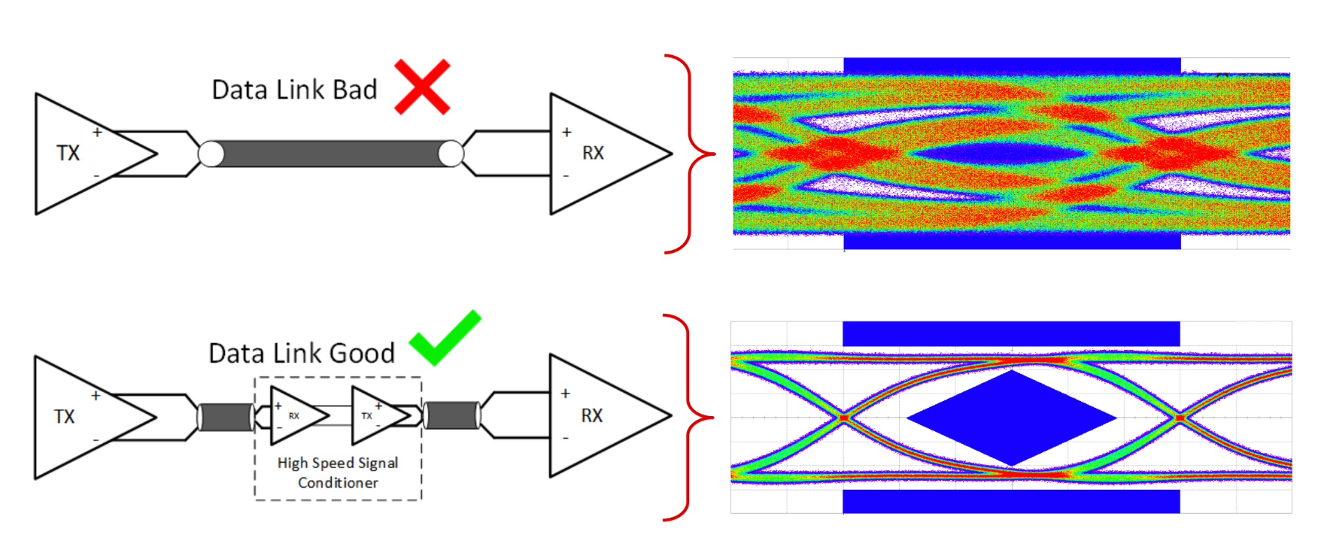}
\label{fig:1a}}
  \hspace{+0.3cm}
\subfloat[]{\includegraphics[width=0.3\textwidth]{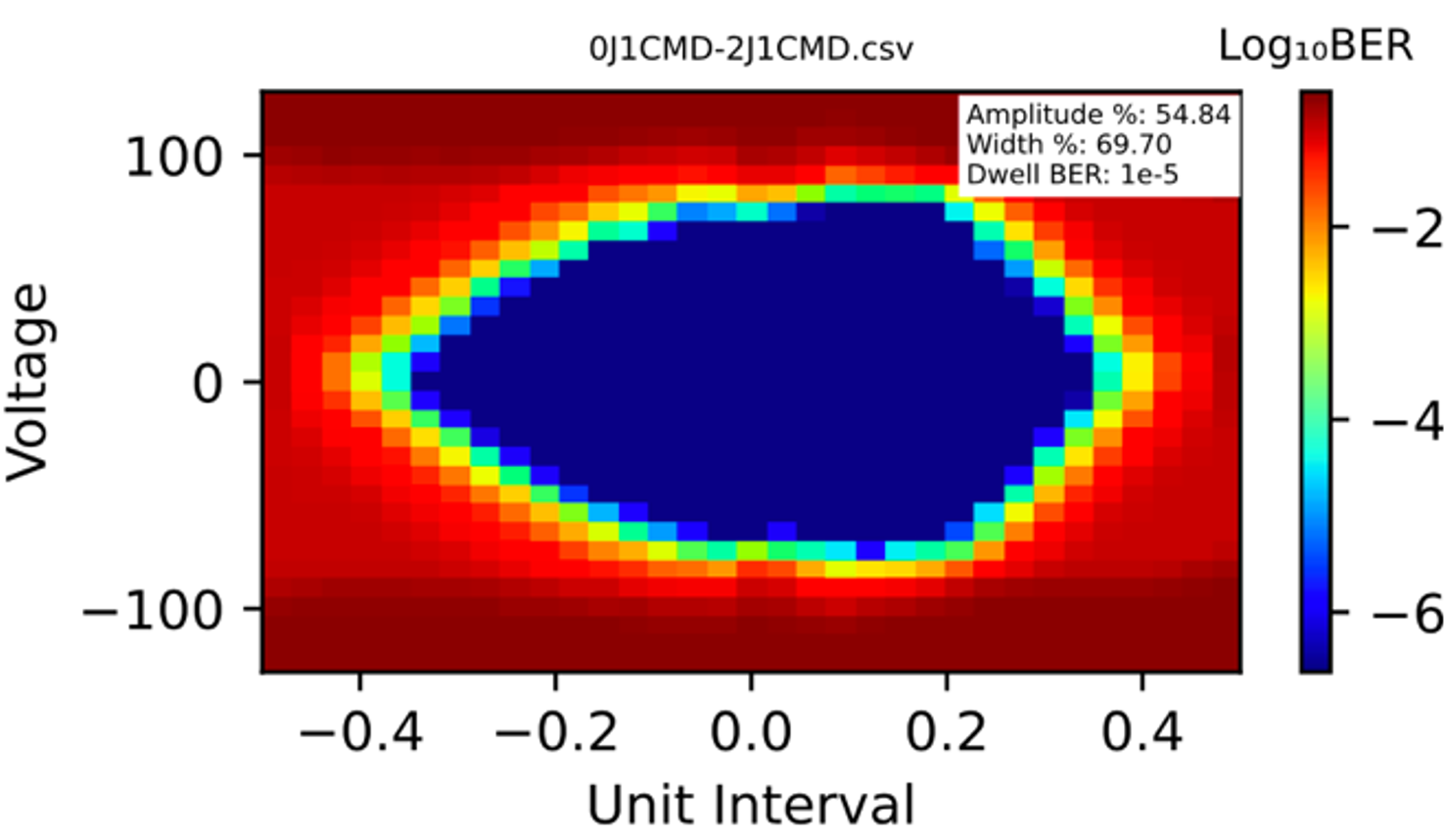}
\label{fig:1b}}

\caption[Short caption]{Examples of (a) a standard eye diagram measured on an oscilloscope (transfer function of the receiver in blue) and (b) a virtual eye diagram obtained from a BER scan over a unit cell.}

\label{fig:1}
\end{figure}

Figure~\ref{fig:1} shows examples of eye and BER scan diagrams, each for a single channel. The BER method allows to automate the measurement of all (up to 32) links in our custom-built assemblies and the scaling of testing multiple assemblies in parallel. Measuring the amplitude and width of the eye opening \cite{2} obtained from the BER test scans offers analogous information to that of standard eye diagrams and allows to judge if the build quality of the links meets the standards in ITk. 

\section{Multi-channel DAQ FPGA-based Framework}

\begin{figure} [htbp]
\centering
\subfloat[]{\includegraphics[width=0.4\textwidth]{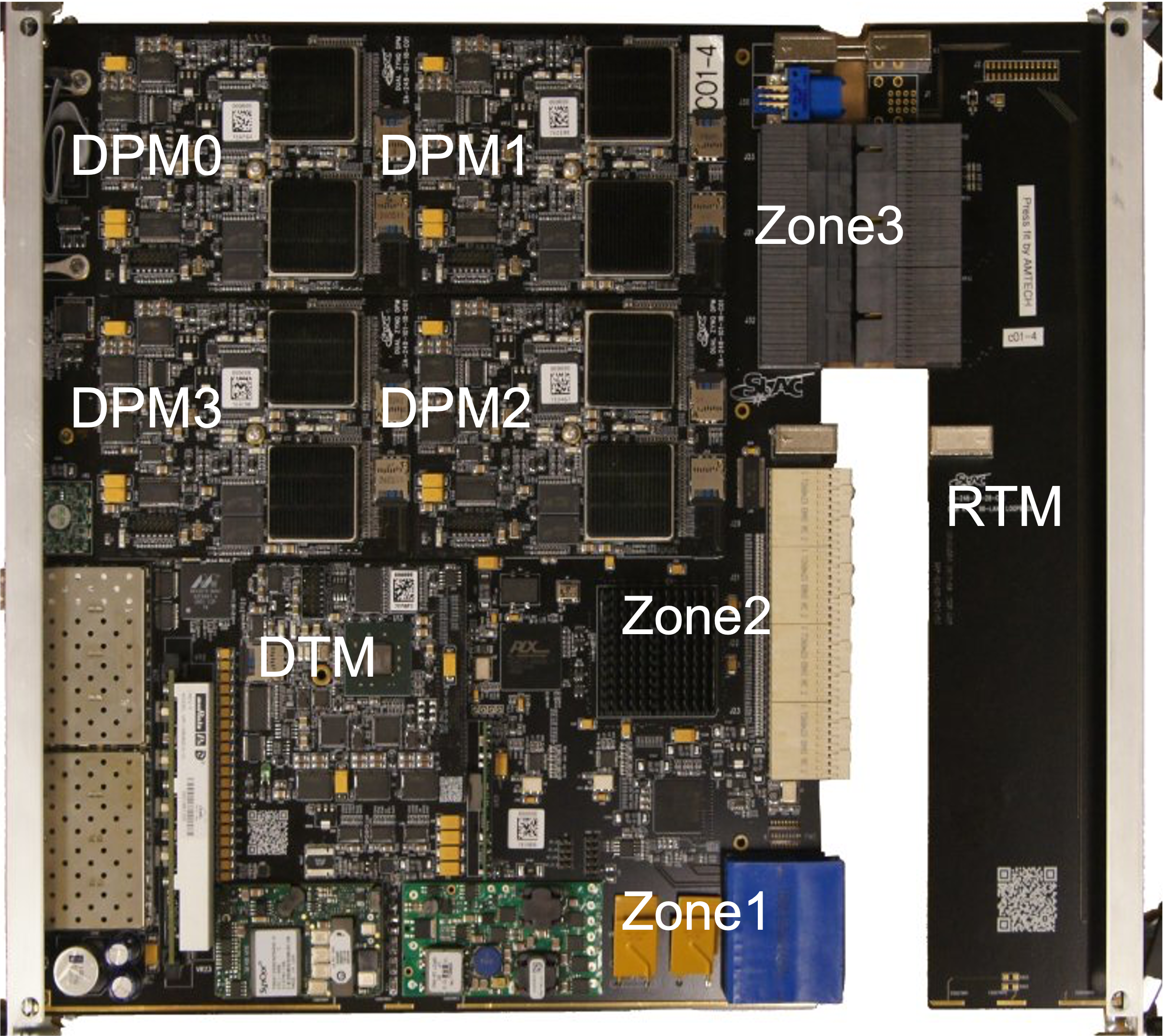}
\label{fig:2a}}
  \hspace{1cm}
\subfloat[]{\includegraphics[width=0.32\textwidth]{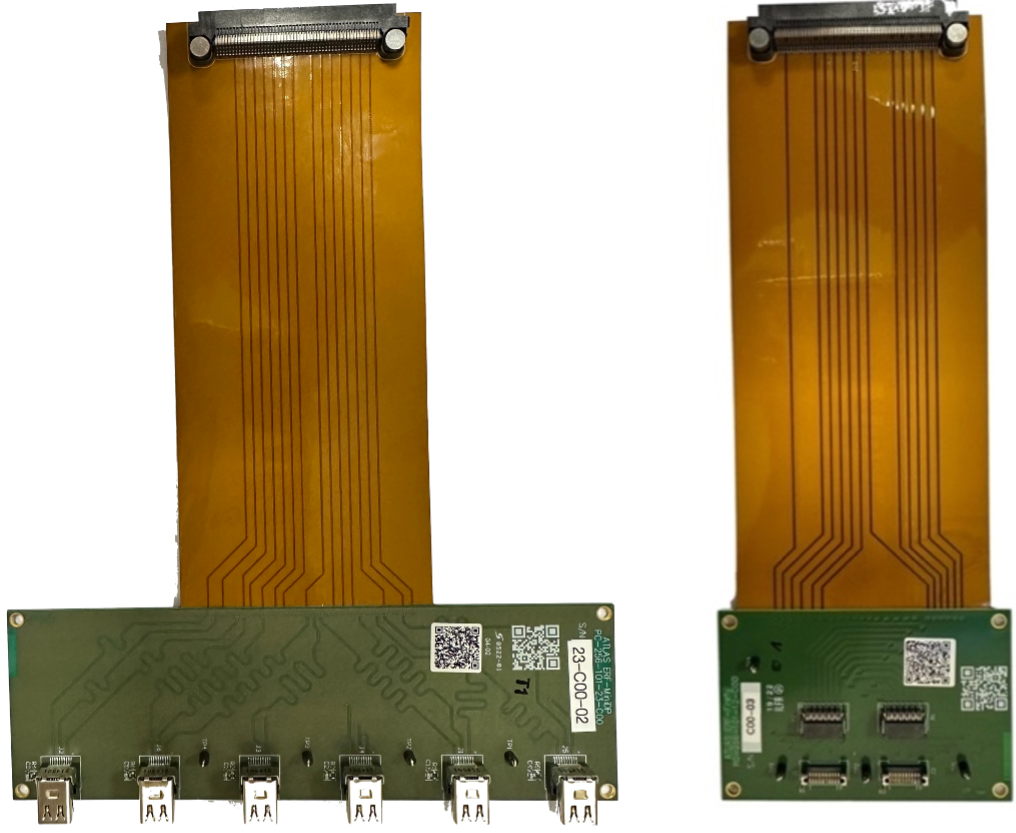}
\label{fig:2b}}
\vspace*{-0.2cm}
\caption[Short caption]{(a) The Cluster on Board (COB), (b) the adapters interfacing further connector types to the RTM, MiniDP (left) and Dual FireFly (right).}

\label{fig:Fig.2}
\end{figure}

The DAQ system for the BER scans is based on the Advanced Telecommunications Computing Architecture (ATCA), having compact shelf and providing scalability to four data link assemblies at the same time. The Cluster on Board (COB), shown in Figure~\ref{fig:2a}, was designed and manufactured by SLAC and provides the framework for the BER test scans. It includes nine Reconfigurable Cluster Elements (RCEs), system-on-chip designs based on the Xilinx Zynq FPGAs, one in the Data Transmission Module (DTM) for synchronization of the electronic equipment \cite{3}, and two each in the four Data Processing Modules (DPM0-3). DPM0-3 provide the capability to send, receive and analyze pseudo-random bit streams at 1.28Gb/s, as well as the programmable biasing of the voltage and time offsets spanning the range of a logic unit. The COB interfaces to the Rear Transmission Module (RTM) and both are powered via the ATCA. For the BER scans, the RTM serves as an interface facilitating communication between the device under tests (DUTs) and DPMs.

\vspace*{-0.4cm}

\begin{figure} [htbp]
\centering
\subfloat[]{\includegraphics[width=0.4\textwidth]{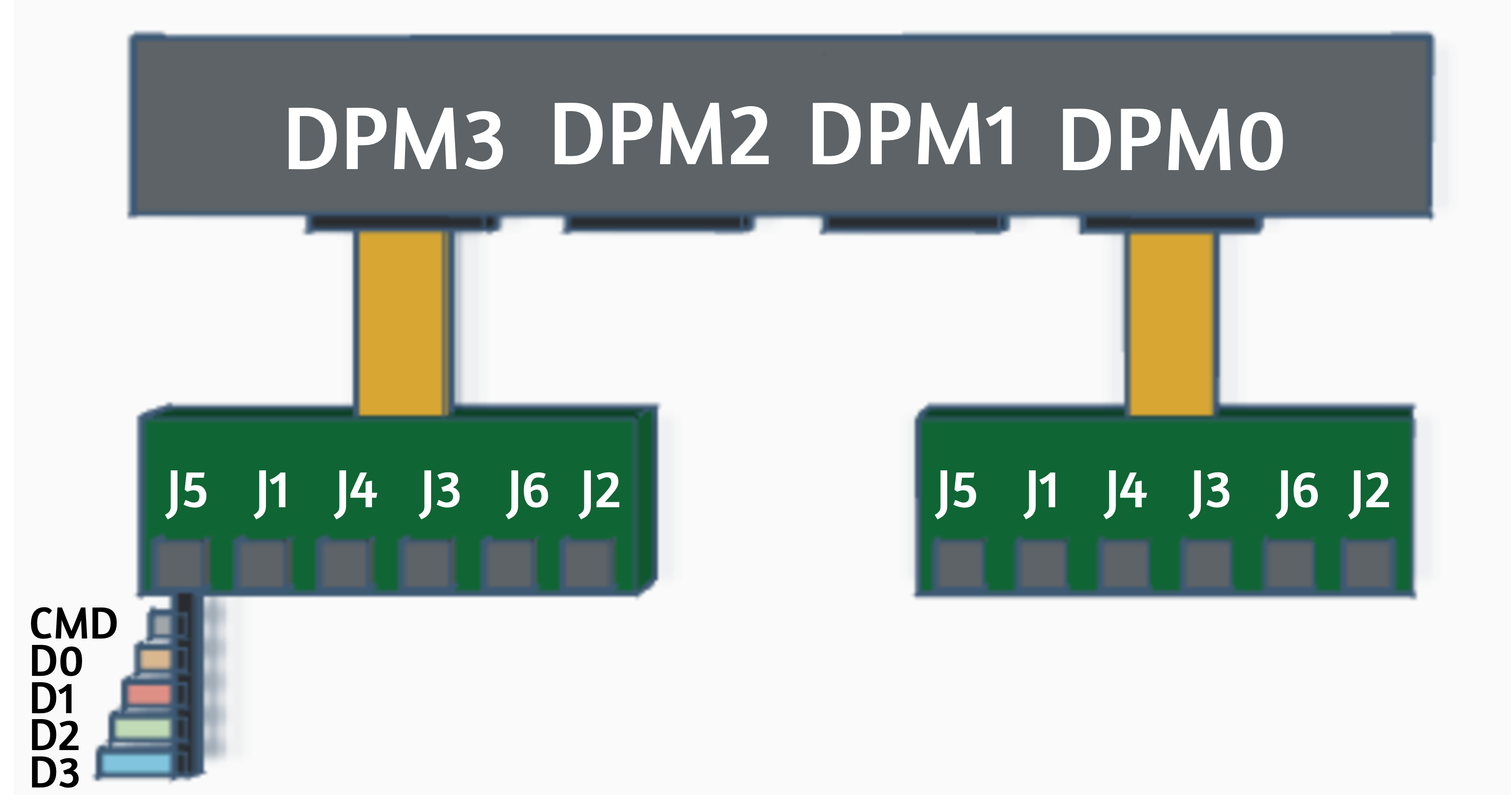}
\label{fig:3a}}
  \hspace{1cm}
\subfloat[]{\includegraphics[width=0.38\textwidth]{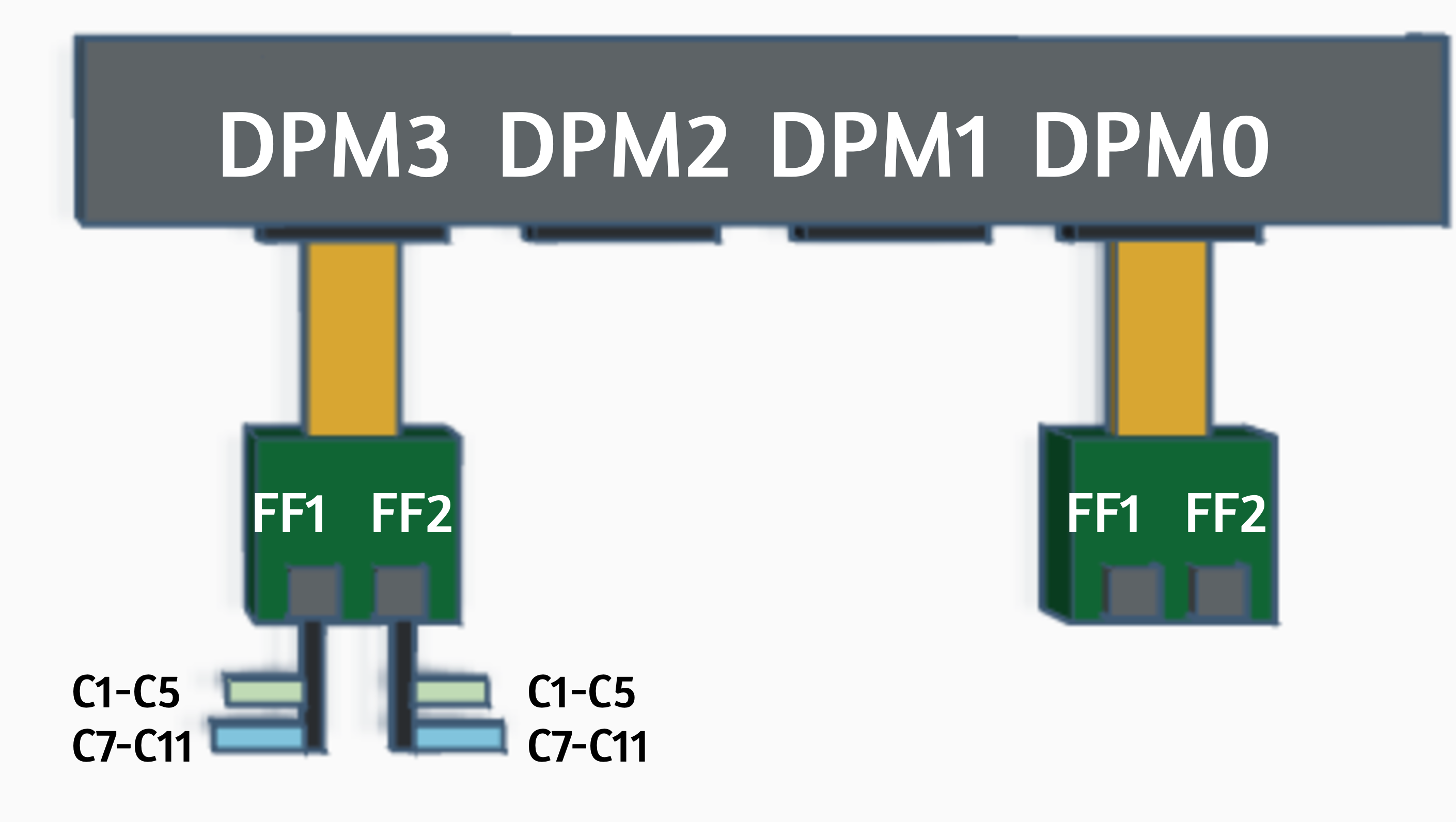}
\label{fig:3b}}
\vspace*{-0.2cm}
\caption[Short caption]{(a) A loop is created for each miniDP connector pair, starting from J1-J1, with six loops (J1-J1 to J6-J6) as DUTs, supporting five differential lanes (CMD, D0-D3) (b) Type-1 Twinax loop with FireFly-1 (FF1-FF1) and FireFly-2 (FF2-FF2) as DUTs, each with 10 lanes (C1-C5, C7-C11).}
\label{fig:Fig.3}
\end{figure}
\vspace*{-0.2cm}
The DAQ system employs two types of connectors: one connects six industry-standard miniDP cables for test system calibration, while the other supports Type-1 Twinax bundle testing, as shown in Figure~\ref{fig:2b} (left and right, respectively). Each connector is integrated through an RTM that supports four channels (DPM0–DPM3) as shown in Figure~\ref{fig:Fig.3}. Two of these channels function as transmitters (TX), while the other two serve as receivers (RX). Transmission loops are established to assess the signal transmission quality of DUTs. Figure~\ref{fig:Fig.3} shows the RTM-to-DUT mapping for both test configurations.

\vspace*{-0.2cm}

\section{Aids in Correlating Virtual Eye Diagrams with Insertion Loss (IL)}

The ITk data links were qualified via S-parameter measurements, with specifications defined by insertion loss (IL). Thus, the outputs of the DAQ system should be correlated against the IL. Additionally, the study \cite{6} demonstrates that IL can be employed for the precise prediction of BER eye amplitude and width. Consequently, the IL of industry-standard miniDP cables was measured using a Vector Network Analyzer (VNA). These cables were then used to generate virtual eye diagrams from BER test scans to calibrate the test system and gain insight into its performance. Different cable lengths, representing varying signal losses, were employed, and all DAQ channels were systematically tested to assess the correlation between the virtual eye openings and the IL.

Ideally, one would use a 4-channel VNA to measure the S-parameters of a differential link. Since we only had a 2-channel VNA available (N9918A FieldFox) we employed the mixed-mode S-parameter approach studied in \cite{4}. It shows a strong correlation between results from a 2-channel VNA and a 4-channel VNA. In this approach, the transmitted signal of a differential link, SDD21, is obtained from multiple measurements with the relation in \eqref{eq:eq-1}. The IL is its complement.

\begin{equation}
\label{eq:eq-1}
SDD21= \frac{1}{2} (S21-S23-S41+S43)
\end{equation}

As an addtional note, the signal-to-noise ratio (SNR) is directly related to the BER \cite{7}. It is influenced by two factors: crosstalk and IL \cite{8}. In our setup, which employs twinax cables, the impact of crosstalk can be disregarded, leaving insertion loss as the sole contributing factor.

\vspace*{-0.2cm}

\section{Results}
 
In this section, the results are presented in two parts, covering the calibration of the test system and the preliminary test results on the first Twinax Type-1 prototype assembly.

\subsection{Calibration of the DAQ System}

To understand channel-to-channel variations in signal transmission within test system, we recorded transmission data using 1m and 5m miniDP cables connected via miniDP adapters between the TX/RX channels. Each adapter provides six miniDP channels (J1-J6), with each channel mapping five RTM differential lanes, as shown in Figure~\ref{fig:3a}. The data lanes have a fixed mapping (CMD-CMD, D3-D0, D2-D1, D1-D2, D0-D3), limiting TX-to-RX connection options. We calibrated the five lanes to evaluate RTM\&COB signal paths, identify high-performing lanes for further tests, and refine the calibration model for transmission variations. This follows to correlating BER scans with the IL of data lane loops in the miniDP channels.

\begin{figure} [ht]
\centering
\subfloat[]{\includegraphics[width=0.326\textwidth]{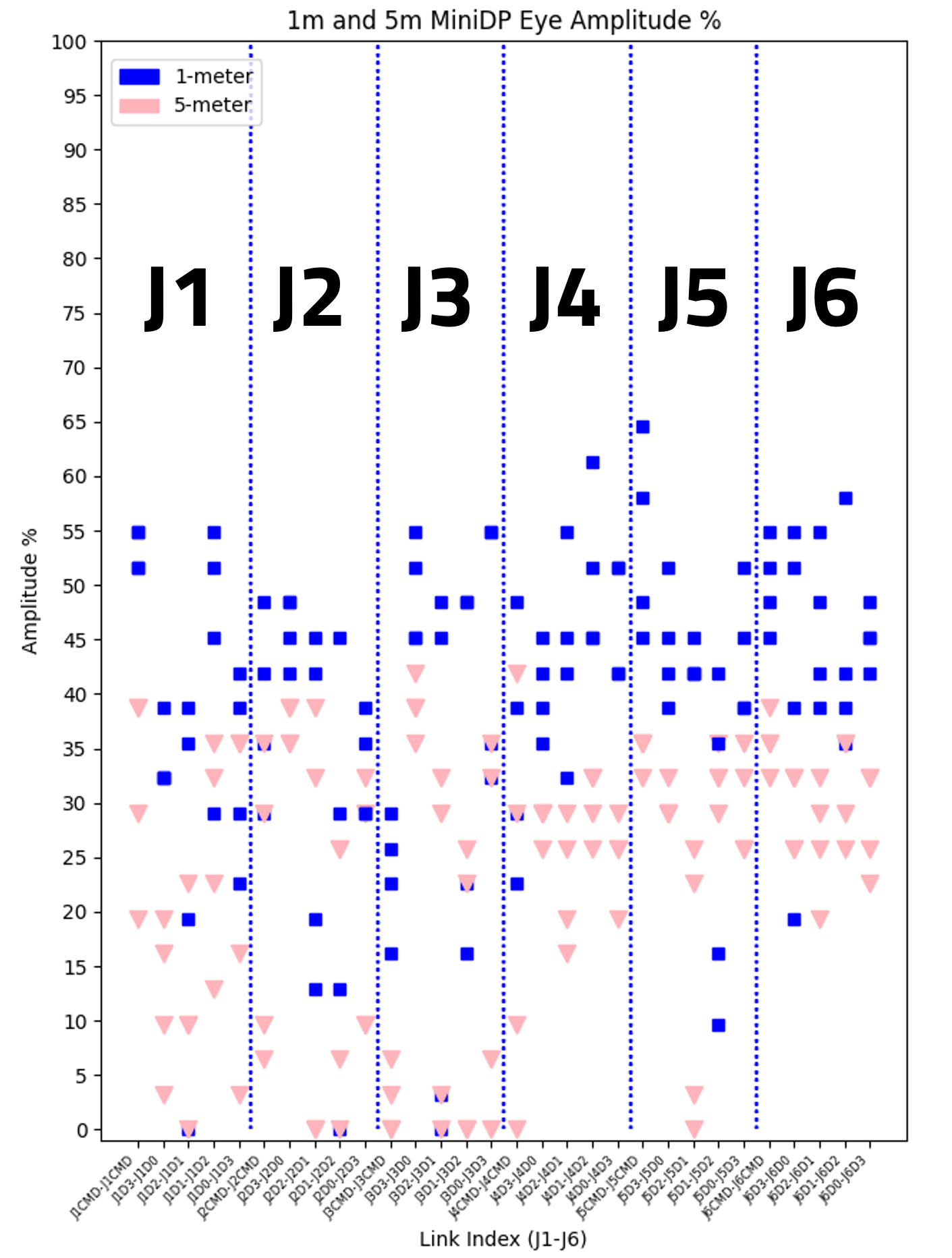}
\label{fig:4a}}
\subfloat[]{\includegraphics[width=0.33\textwidth]{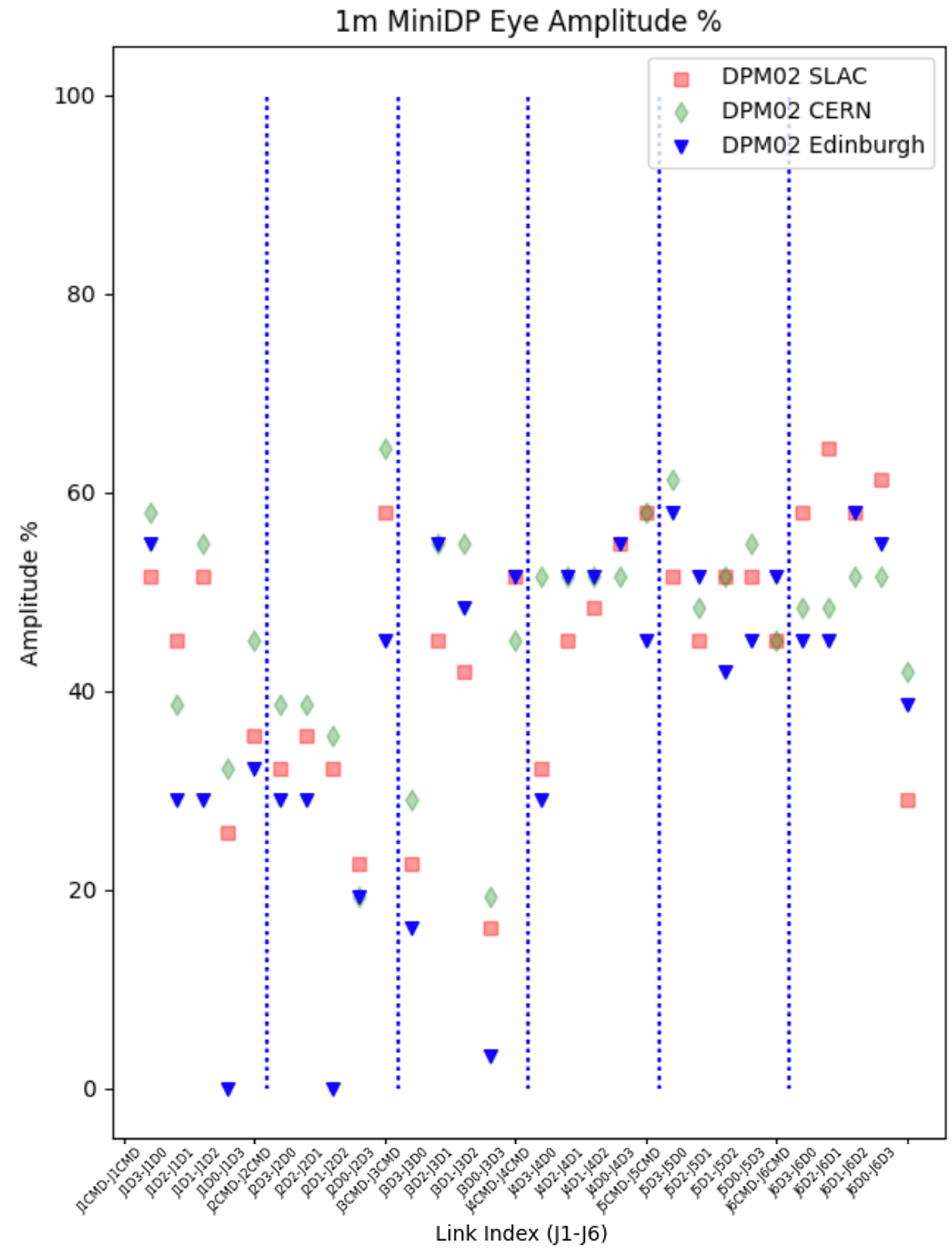}
\label{fig:4b}}
\vspace*{-0.2cm}
\caption[Short caption]{The amplitudes of BER diagrams for 1m\&5m miniDPs, (b) the miniDP test outputs from three test centers. Js represent miniDP loops, with five mappings explained above per loop.}
\label{fig:Fig.4}
\end{figure}
\vspace*{-0.056cm}
Sharper signal edges indicate higher transmission quality, which corresponds to higher amplitude values on the BER diagrams \cite{5}. Therefore, the primary focus will be on the amplitudes of BER scans. Figure~\ref{fig:4a} shows the amplitude rates of the multiple BER scans (the different DUT configurations) recorded for 1m and 5m miniDP cables. Compared to 1m cables a $\sim$15\% reduction in amplitude is observed for 5m cables. Meanwhile, Figure~\ref{fig:4b} compares the amplitude rates for 1m miniDP cables recorded with three independent setups, at SLAC, CERN and Edinburgh. The results correlate well and show that the channels J4-J6 yield the best transmission properties, with amplitude levels between 40-60\%. These channels will be prioritized for QC testing, while it is suspected that the lower performance of channels J1-J3 may be due to design flaws in the RTM. 

\begin{figure}[htbp]
\centering 
\includegraphics[width=.5\textwidth,clip]{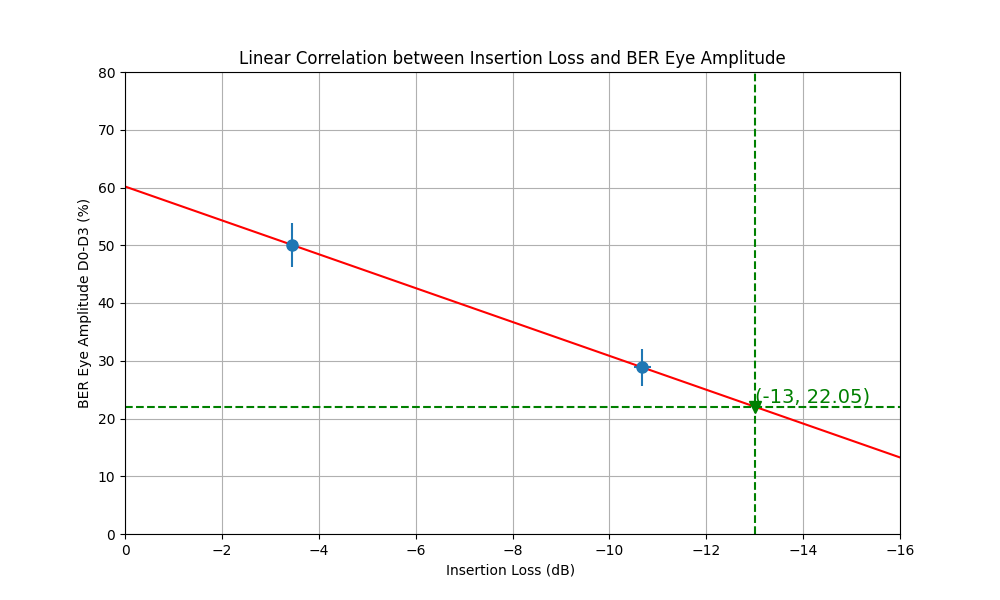}
\vspace*{-0.2cm}
\caption{\label{fig:5} The linear correlation between IL and BER scan amplitude for 1m\&5m miniDP loops.}
\end{figure}

Figure~\ref{fig:5} illustrates the linear relationship between IL distributions and eye amplitude rates from BER scans at 1.28GHz, carried out CMD and D0-D3 lanes using 1m and 5m cables. Accurate signal reconstruction requires a sampling frequency $f_s$ at least twice the highest signal frequency, as defined by $f_s\geq 2f_{max}$, where $f_{max}$ (the Nyquist frequency \cite{9}) is 640 MHz. The results showed consistent IL values of $-3.5\pm0.11$dB and $-10.8\pm0.17$dB for 1m and 5m, respectively. Eye opening data of the same miniDP cables tested in channels J4-J6 yield amplitudes of $50.0\%\pm3.9$ and $28.9\%\pm3.2$. In the plot, the intersection point (at -13dB, 22.05\%) highlights the allocated ITk signal loss budget. The system meets the pass criteria when the BER eye amplitude rate exceeds 22.05\%. This value reflects the maximum allowed IL while maintaining the minimum required BER eye amplitude, ensuring reliable operation within specified limits. This result can help to define the pass/fail criteria for data links during production QC.

\vspace*{-0.2cm}
\subsection{The First Type-1 Twinax Bundle Prototype}

\begin{figure} [htbp]
\centering
\subfloat[]{\includegraphics[width=0.25\textwidth]{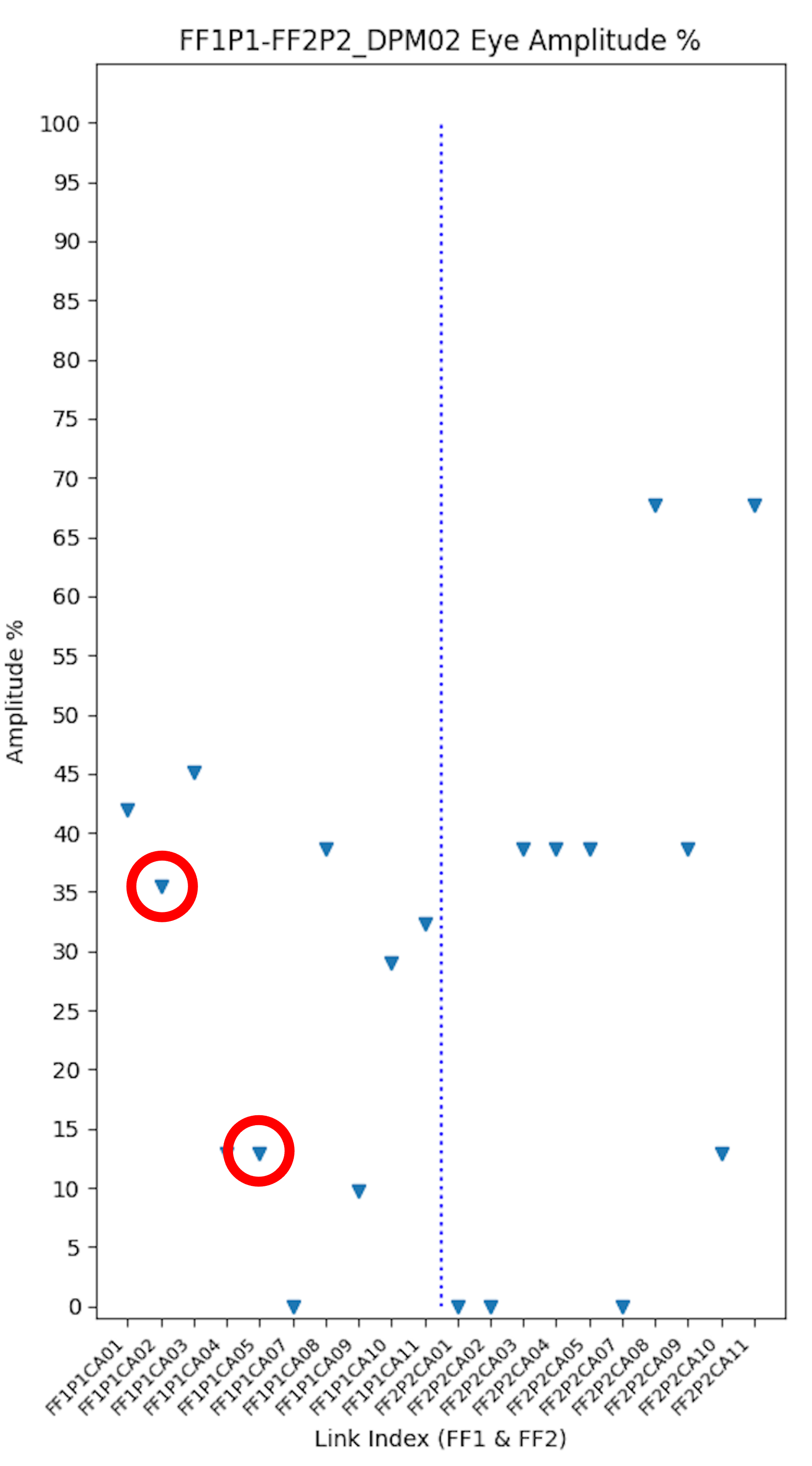}
\label{fig:6a}}
  \hspace{+0.3cm}
\raisebox{0.7cm}{\subfloat[]{\includegraphics[width=0.5\textwidth]{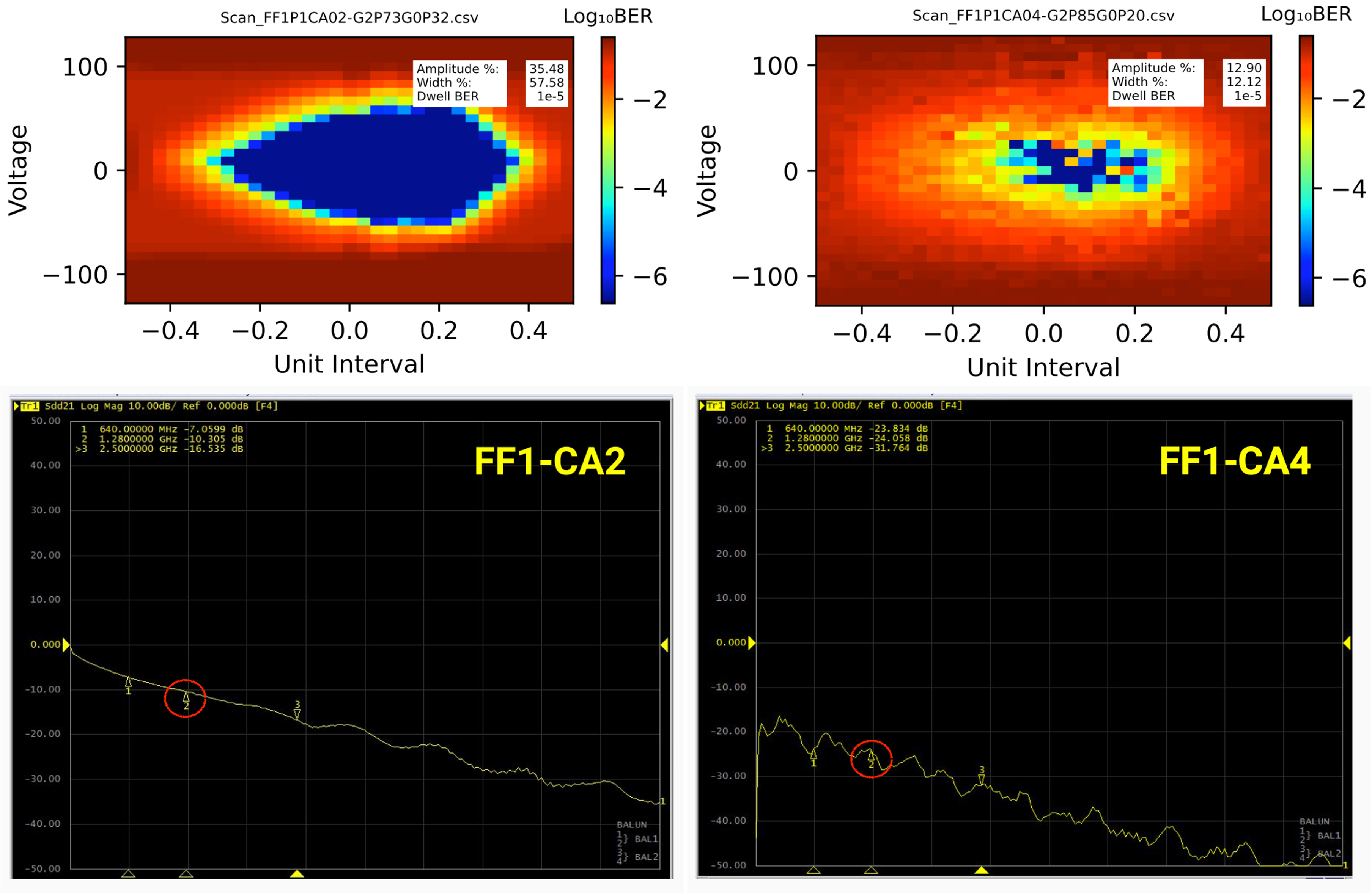}
\label{fig:6b}}}
\vspace*{-0.2cm}
\caption[Short caption]{(a) BER diagram amplitudes for channels in the first Twinax prototype, (b) BER diagrams and VNA IL for channels with strong and weak signal transmissions.}
\label{fig:Fig.6}
\end{figure}

A first prototype data link assembly was tested in Edinburgh, featuring two ribbons, 304cm and 315cm long, each with 10 Twinax links. Firefly connectors FF1\&FF2, with their pads CA1-CA5 and CA7-CA11 (see in Figure~\ref{fig:3b}), terminated the two ribbons on the side opposite to the common connector on RTM. Figure~\ref{fig:6a} shows the eye opening amplitudes for 20 data links (10 per ribbon, separated by a dotted line), indicating that 8 links failed to meet the performance threshold, with BER scan amplitudes below the $\sim$22\% criterion established in Figure~\ref{fig:5}.

Figure~\ref{fig:6b} focuses on VNA measurement and virtual eye diagram of two channels FF1-CA2 and FF1-CA4, as examples of strong and weak data transmissions. The top shows the virtual eye diagrams, clearly distinguishing the two cases, measuring amplitudes of 35.5\% and 12.1\%, respectively. Below them the VNA IL measurements of these two channels are shown, yielding ILs of -7.1dB and -23.8dB at 640MHz and -10.3dB and -24.1dB at 1.28GHz. Only the first meets the signal loss budget of the ITk data links. This result matches the findings from the BER scan, confirming the consistency of the measurements. Finally, gathering data from various test centers and DUTs, along with comparing measurements, may help determine appropriate pass/fail thresholds for amplitude in virtual eye diagrams for further steps.
\vspace*{-0.47cm}
\section{Conclusion and Further Plans}
\vspace*{-0.15cm}
We developed a DAQ system to record virtual eye diagrams from BER scans to enable automated QC testing of large numbers of custom data links. These systems were built independently and yield consistent results in a cross-check using industry standard links. Correlating VNA IL data with the virtual eye diagrams has been started and will inform the pass/fail criteria for the QC tests. A first prototype ITk data assembly has been evaluated on the system at Edinburgh.

\end{document}